\documentclass{aastex631}
\usepackage{CJK}
\usepackage{soul}
\usepackage{multirow}

\shorttitle{A Versatile VLBI Digital Backend}
\shortauthors{Jiyun Li, Renjie Zhu, Shaoguang Guo et al.}
\graphicspath{{./}{figures/}}
\graphicspath{{../}{../}}
\begin{document}
\begin{CJK*}{UTF8}{gbsn}

\title{Development and Performance Validation of a Versatile VLBI Digital Backend Using the ROACH2 Platform}

\correspondingauthor{Shaoguang Guo}
\email{sgguo@shao.ac.cn}

\author[0009-0006-0056-7199]{Ji-Yun Li(李纪云)}
\affiliation{Shanghai Astronomical Observatory,  Chinese Academy of Sciences, Shanghai 200030, China;}
\affiliation{Shanghai Key Laboratory of Space Navigation and Positioning Technology, Shanghai 200030, China}

\author{Renjie Zhu(朱人杰)}
\affiliation{Shanghai Astronomical Observatory,  Chinese Academy of Sciences, Shanghai 200030, China;}
\affiliation{Shanghai Key Laboratory of Space Navigation and Positioning Technology, Shanghai 200030, China}
\affiliation{University of Chinese Academy of Sciences, Beijing 100049, China}
\affiliation{ Key Laboratory of Radio Astronomy and Technology, Chinese Academy of Sciences, Beijing 100101, China}

\author[0000-0003-0181-7656]{Shaoguang Guo(郭绍光)}
\affiliation{Shanghai Astronomical Observatory,  Chinese Academy of Sciences, Shanghai 200030, China;}
\affiliation{Shanghai Key Laboratory of Space Navigation and Positioning Technology, Shanghai 200030, China}
\affiliation{University of Chinese Academy of Sciences, Beijing 100049, China}
\affiliation{ Key Laboratory of Radio Astronomy and Technology, Chinese Academy of Sciences, Beijing 100101, China}

\author{Ping Rui(芮萍)}
\affiliation{Shanghai Astronomical Observatory,  Chinese Academy of Sciences, Shanghai 200030, China;}
\affiliation{Shanghai Key Laboratory of Space Navigation and Positioning Technology, Shanghai 200030, China}

\author[0000-0003-4853-7619]{Zhijun Xu(徐志骏)}
\affiliation{Shanghai Astronomical Observatory,  Chinese Academy of Sciences, Shanghai 200030, China;}

%\collaboration{6}{(AAS Journals Data Editors)}



\begin{abstract}

Customized digital backends for Very Long Baseline Interferometry (VLBI) are critical components for radio astronomy observatories. There are several serialized products such as the Digital Baseband Converter (DBBC), Reconfigurable Open Architecture Computing Hardware (ROACH) Digital BackEnd  (RDBE), and Chinese Data Acquisition System (CDAS). 
However, the reliance on high-speed analog-to-digital converters (ADC) and Field Programmable Gate Arrays (FPGAs) often necessitates dedicated hardware platforms with long development cycles and prohibitive cost,
limiting scalability and adaptability to evolving observational needs. 
To address these challenges, we propose a design leveraging the versatile and cost-effective ROACH2 hardware platform, developed by CASPER (Collaboration for Astronomy Signal Processing and Electronics Research).
ROACH2's mature technology and streamlined firmware development capabilities significantly 
reduce the hardware platform's development cycle and cost,
making it ideal for modern astronomical applications.
This VLBI digital backend, based on the ROACH2 platform, incorporates key technologies such as Polyphase Filter Banks (PFB) algorithm implementation,  digital complex-to-real baseband signal conversion, Mark5B data formatter design and two-bit optimal threshold quantization.
These features ensure compatibility with existing systems while providing enhanced performance.
The backend's performance was validated through multi-station VLBI experiments,
demonstrating its ability to achieve good correlation fringes compared to the customized CDAS2-D system.
Furthermore, this platform offers flexibility for rapid deployment of additional digital backends,
such as those for spectral line observations,
showcasing its potential for broader astronomical applications.

\end{abstract}



\section{Introduction} \label{sec:intro}

\label{sect:intro}

In recent years, there has been a remarkable surge in the development of large-aperture single-dish antenna and aperture array telescopes, aimed at significantly improving resolution and sensitivity. 
This growing roster includes notable telescopes such as 
the Five-hundred-meter Aperture Spherical Radio Telescope (FAST) \citep{Nan+etal+2011, Li+etal+2018}, 
the 110 m QiTai Radio Telescope (QTT) \citep{wangna-qtt}, 
the Tianma 65m Radio Telescope \citep{liu2024tianma}, 
the Atacama Large Millimeter Array (ALMA) \citep{Baudry+2008}, 
FAST Core Array \citep{jiangpeng-fast-core-array},
and the Square Kilometre Array (SKA) \citep{Dewdney+etal+2009}. 
These facilities exemplify the global drive to address critical questions in astrophysics, 
ranging from the structure of the universe to the origins of celestial phenomena.

The advent of such cutting-edge telescopes has elevated the demands on associated instrumentation,
particularly digital backends, which are integral to acquiring and processing high precision, wide bandwidth signals.
Radio astronomy digital backends are designed to perform tasks such as analog-to-digital conversion,
baseband signal processing, data formatting, data transfer and local monitoring.
These systems cater to various scientific objectives and can be categorized into  
VLBI digital terminals \citep{Thompson+etal+2017},
pulsar digital terminals \citep{Xu+etal+2015}, 
and spectral line digital terminals \citep{Wu+etal+2017}.

The Shanghai Astronomical Observatory (SHAO) of the Chinese Academy of Sciences has been 
a pioneer in the research and development of VLBI digital backends.
Over the years, SHAO has successfully upgraded its systems,
culminating in the second-generation Chinese Data Acquisition System Generation (CDAS2) \citep{Zhu+etal+2016}
and is currently making efforts to develop the third-generation one  (CDAS3). 
The first generation VLBI data acquisition system was the Analog Baseband Converter (ABBC) \citep{Xiang+2005}, 
which converts the IF signal to baseband in the analog domain while the  ADC chip samples the baseband signal directly and sends this baseband data to the FPGA for further processing,
CDAS was employed during the initial phase of the China Lunar Exploration Program (CLEP) \citep{Li+etal+2019}.
However, the rapid pace of technological advancements rendered some components  obsolete, highlighting the challenges of maintaining such hardware-intensive systems.
As the second phase of the lunar exploration project commenced,
the VLBI data acquisition system was upgraded by SHAO, 
which brought about the creation of the first generation of VLBI digital baseband conversion system (CDAS).
Unlike the ABBC design, CDAS utilizes a high-speed ADC and FPGA architecture, 
placing the sampling module closer to the RF to facilitate FPGA firmware upgrades. 
Despite these improvements, earlier platforms faced challenges such as limited FPGA resources
and complex system designs requiring cascading multiple FPGA chips for processing,
which complicated maintenance and scalability.
%However, due to the limited FPGA programmable resources at that time, 
%the CDAS hardware platform demanded cascading multiple FPGA chips for data processing, which led to a complex system design and presented maintenance challenges.
Notwithstanding, today's abundance of FPGA programmable resources has streamlined the design of the VLBI digital terminal system.
Indeed, this technological advancement is evident in the CDAS2 platform, which not only embodies this feature but was also successfully deployed in the third phase of the lunar exploration mission.

Widely adopted VLBI digital backends include 
the Digital BaseBand Converter (DBBC) series \citep{Tuccari+etal+2012}, % of the European VLBI Network (EVN), 
the ROACH Digital BackEnd (RDBE) series \citep{Niell+etal+2010} based on ROACH in the United States, 
the Analog to Digital Sampler (ADS) series \citep{Takefuji+etal+2010} in Japan, 
and the Chinese CDAS series. 
The development cycle of an astronomical digital backend typically includes stages such as schematic design, printed circuit board design, board testing, embedded control system design, FPGA interface design, FPGA signal processing algorithm design, and so on. 
The time and cost associated with hardware testing of circuit boards, embedded system development, 
and hardware interface design are significant for custom-made of astronomical digital backends. 
%Consequently, the upgrade cycle of astronomical digital backends lags behind the rising demands of astronomical observation and struggles to fulfill longstanding needs of radio astronomy observations,which poses major impediments to the progression of radio astronomy science. 
Such bespoke designs of astronomical digital backends entail long development cycles and significant costs,
impeding their ability to meet the fast-evolving demands of modern astronomical observations.
Moreover, their limited production volumes exacerbate manufacturing costs, further restricting accessibility and scalability. 
Due to their specialized nature, astronomical digital backends have limited production volumes, resulting in significantly higher manufacturing costs.
Considering the enduring needs for astronomical digital terminals, 
utilizing established commercial platforms for further development can be advantageous, as it would stimulate the advancement of these backends.
Such a commercial platform should provide  a mature embedded control system, reliable hardware circuit design, 
and iterative hardware upgrades, which can, to an extent,
reduce the design cost of astronomical digital backends.
This approach can shorten the development cycle and  speed up the deployment for scientific applications.
Designers of astronomical digital terminals would thus need to focus primarily  on developing FPGA data processing algorithms on commercial systems. 
By implementing different signal processing algorithms on the same FPGA platform,
it is possible to develop multifunctional digital backends.
This approach allows for seamless adaptation across diverse observational modes, including VLBI, pulsar \citep{Pei+etal+2017}, and spectral line observations,
thereby maximizing operational efficiency and scientific output.
Such a strategy would dramatically boost work effectiveness for observers and enhance the utilization rate of the digital backends.

The work presented in this paper introduces a VLBI digital backend on the ROACH2 platform, 
highlighting its innovative use in reducing design complexity and cost.
By employing key technologies such as PFB algorithms, 
real-to-complex signal conversion, and Mark5B data formatting, 
the system addresses the demands of high-precision astronomical observations while enabling rapid customization for various scientific missions.

\section{VLBI Digital Backend Based on ROACH2}
%\label{sect:Obs}

ROACH2\footnote{\url{https://casper.astro.berkeley.edu/wiki/ROACH}} is an advanced radio astronomy signal processing platform equipped with a robust Linux embedded control system,
enabling efficient FPGA secondary development and a rich set of data processing  Intellectual Property (IP) cores. 
Leveraging the capabilities of the ROACH2 platform, designers can develop various backends tailored to observational  needs,
such as VLBI digital backends, pulsar digital backends, spectral line digital backends.
This section provides a detailed overview of the VLBI digital backends,
emphasizing its design, architecture, and algorithmic innovations.

\subsection{VLBI Digital Backend Overview}

\begin{figure}
   \centering
   \includegraphics[width=\textwidth, angle=0]{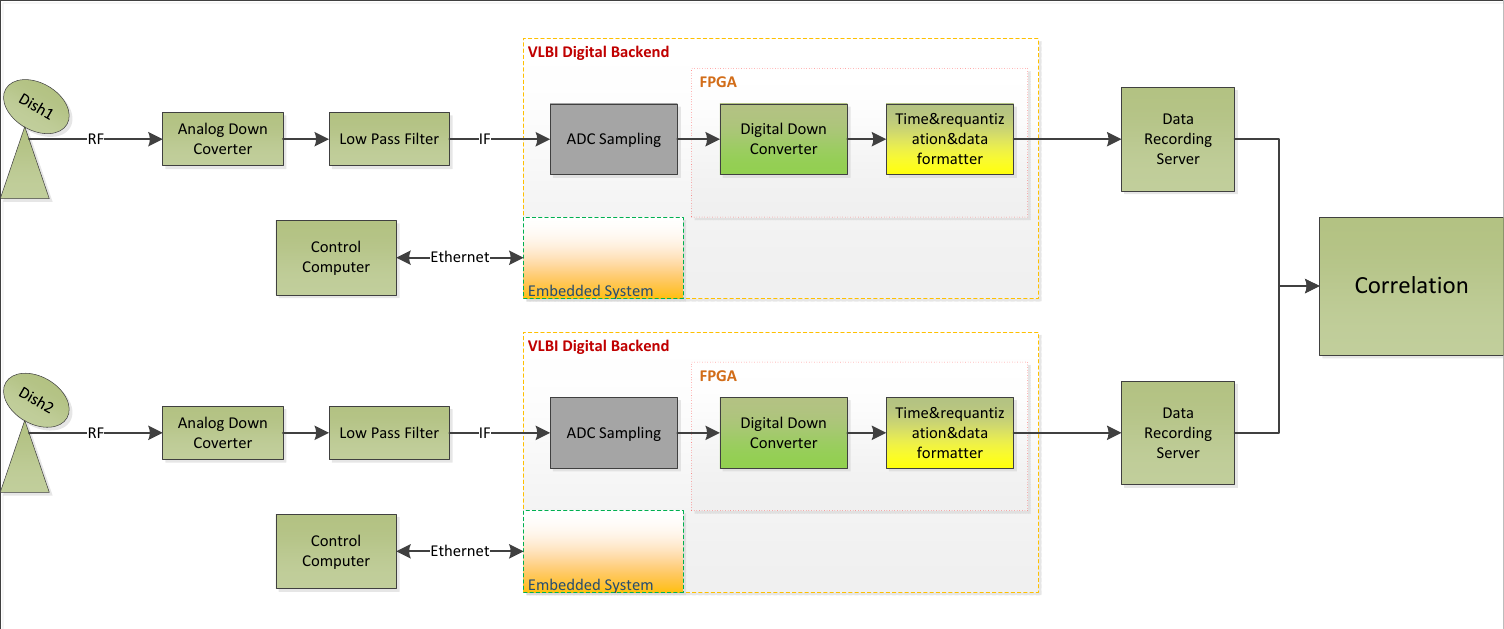}
   \caption{Block diagram of the VLBI digital backend system, illustrating the signal flow from the radio frequency (RF) output of the antennas (Dish1 and Dish2) through the ADC, low-pass filters, and ADC sampling units. The digital down-converter, implemented in an FPGA, performs signal processing, including time and re-quantization, followed by data formatting. The processed data is transmitted to the data record server via an embedded system and Ethernet, with subsequent correlation for further analysis.}
   \label{Fig1}
   \end{figure}

Figure \ref{Fig1} illustrates the schematic design of the VLBI digital backend system, 
highlighting its modular architecture,
a fundamental piece of equipment in the VLBI data acquisition system. 
The backend performs the crucial functions such as digital baseband conversion  for cross-correlation operations across multiple radio telescope stations.
Presently, an integration of high-speed ADC and FPGA architecture is being employed universally in all prominent VLBI digital backends to proficiently preprocess the wideband analog IF signal. 
The VLBI digital backend comprises primarily  of an embedded control system, a high-speed analog-to-digital converter (ADC), and a robust FPGA for data preprocessing at high speeds.
The high-speed ADC module converts the wideband analogue IF signal into a wideband digital IF signal and sends it to the FPGA for preprocessing.

%      A figure as large as the width of the column
%-------------------------------------------------------------
%
%      One column rotated figure
%-------------------------------------------------------------

High-speed data preprocessing within the FPGA effectively converts the wideband digital IF signal into multiple digital baseband signals with equal bandwidths.
These digital baseband signals are then quantized to two bits,
and transmitted through the 10GbE network.

The embedded control system, 
serving as the backbone of the VLBI digital backend, 
manages critical tasks such as configuring the ADC chip, exchanging data with the FPGA, 
and maintaining seamless communication with the field system (FS) \footnote{\url{https://github.com/nvi-inc/fs}} control computer, among other responsibilities.
This essential component ensures the smooth and efficient operation of the system,
enabling precise and reliable VLBI observations.

\subsection{ROACH2 Platform Introduction}

\begin{figure}
   \centering
   \includegraphics[width=\textwidth, angle=0]{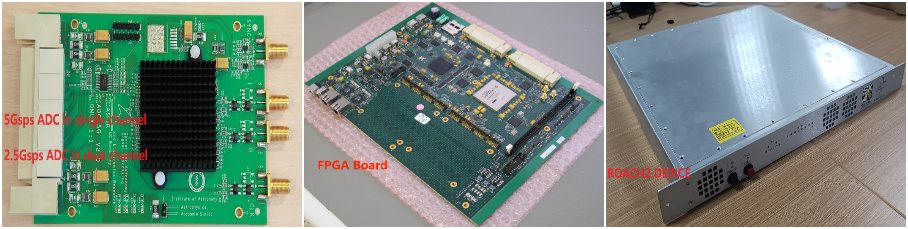}
   \caption{Key components of VLBI digital backend system: (Left) High performance  ADC module capable of 5Gsps in single-channel mode or 2.5 Gsps in dual-channel mode; (Center) FPGA board for digital signal processing and real-time computation; (Right)ROACH2 device for high-speed data processing}
   \label{Fig2}
   \end{figure}

The Collaboration for Astronomy Signal Processing and Electronics Research  (CASPER) in the United States  developed the ROACH2 platform, as shown in Figure \ref{Fig2}.
This platform is widely adopted in radio astronomy signal processing, particularly for applications such as beamforming, real-time correlators, spectrometers, and pulsar detection systems, enabling high-performance and flexible instrumentation in various radio telescope facilities.

CASPER maintains its position at the forefront of innovation by continuously developing and introducing new platforms for radio astronomy signal processing. Their comprehensive portfolio includes the Square Kilometer Array Reconfiguration Application Board (SKARAB), Smart Network ADC Processor (SNAP), ROACH1, and ROACH2. 
Each platform is designed to meet the diverse and unique requirement of radio astronomy digital backends.

Among this suite of platform, ROACH2 has emerged as the preeminent hardware solution, garnering widespread adaption across the astronomical community.
Its versatility is demonstrated by its extensive application across various observations, including SKA multibeam synthesis, spectral line observation, pulsar observation, VLBI observation, and so on.
ROACH2 hardware platform provides significant advantage to designers 
through its wide range of interfaces.
With these rich resources, designers are relieved of routine implementation tasks,
enabling them to focus their efforts on developing fundamental algorithms.
Thus, the ROACH2 platform is not only a powerful tool 
but also a transformative enabler for designers in radio astronomy and other fields,
fostering innovation and advancing scientific exploration.

%      A figure as large as the width of the column
%-------------------------------------------------------------
%
%      One column rotated figure
%-------------------------------------------------------------

The ROACH2 platform, distinguished by its exceptional computing capabilities and versatility in astronomical applications, 
incorporates a PowerPC405 processor with a Linux-based embedded control system to streamline the development of the astronomical digital backend.
Due to  storage constraints of the embedded control system,
storing multiple FPGA configuration files may prove impossible.
To address this limitation, Network File System (NFS) is utilized, 
providing a means of sharing FPGA configuration files via a TCP/IP (Internet Protocol) network with the embedded control system.
With the assistance of the serial debugging tool minicom, 
the Linux operating system supports setting up NFS, IP (Internet Protocol), DHCP services, 
and other interface characteristics for the embedded control system. 

The embedded control system orchestrates data exchange with the FPGA, 
establishes an Ethernet connection to the host computer, 
and initiates the FPGA loading process. 
This system is particularly useful in standard VLBI observation scenarios, 
where the host computer delivers commands via the KATCP (Karoo Array Telescope Control Protocol)  \footnote{\url{https://casper.astro.berkeley.edu/wiki/KATCP}}to the embedded control system.
The system parses these commands, 
configures FPGA shared registers accordingly,
and provides real-time status updates to the host computer.

The ROACH2 hardware platform hosts two high-speed ADC cards, which convert analog IF signals into digital IF signals. 
Each ADC card supports single-channel and dual-channel operation in 8-bit sampling modes. 
In single channel mode, 
the maximum sampling rate is 5Gbps, with an observation bandwidth of 2.5GHz. 
The dual-channel configuration operates at 2.5Gbps per channel,
providing a bandwidth of 1.25GHz for each channel.

The platform's eight 10GbE ports facilitate high-speed data transmission,
theoretically handling a maximum data transfer rate of up to 80Gbps.
The ROACH2 platform uses a Xilinx Virtex6sx475t FPGA (476k Logic Cells, 2016 DSP Slices, 38MB RAM) for real-time digital signal processing and backend data handling.
Complementing this, CASPER provides a range of IP cores,
tailored for processing astronomical data, 
such as FFT IP cores, polyphase filter bank IP cores, 
and 10GbE IP cores, etc.

\section{VLBI Digital Backend Design}

As a versatile and robust radio astronomy signal processing platform, 
the ROACH2 platform provides 
ADC interfaces, 10GbE interfaces, and integrated control interfaces.
This allows designers to focus on implementing signal processing algorithms. 
This section presents an introduction to the overarching framework architecture and highlights key design elements of the VLBI digital backend based on the ROACH2 platform.

\subsection{Overall System Architecture}

\begin{figure}
   \centering
   \includegraphics[width=\textwidth, angle=0]{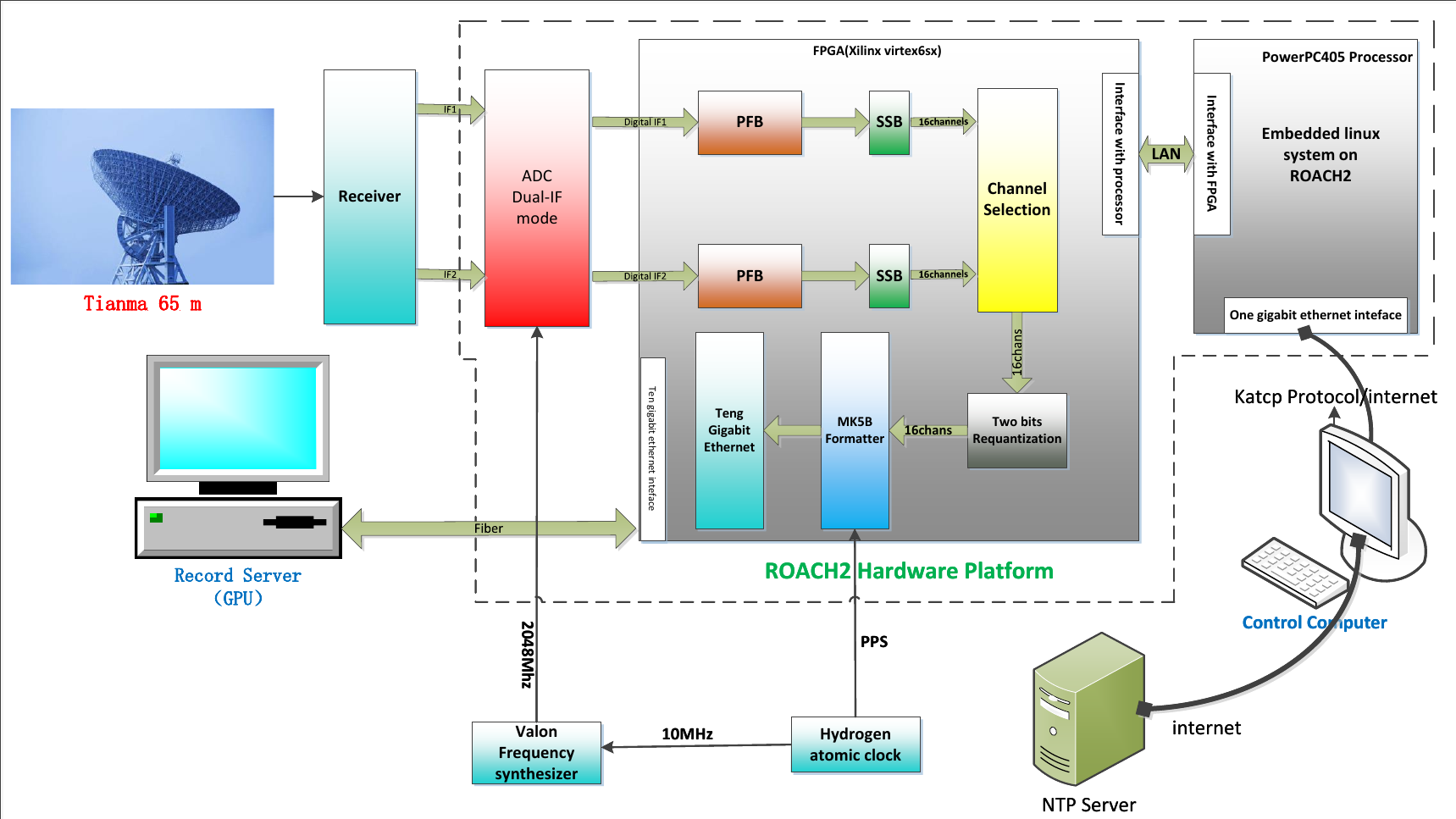}
   \caption{System architecture of the Tianma 65m VLBI station utilizing the ROACH2 hardware platform. The receiver, followed by ADC in dual-IF mode and PFB processing. Subsequent steps include Sideband Separation (SSB), channel selection, and data formatting using the MK5B formatter with two-bit re-quantization, synchronized with the hydrogen maser clock via PPS and 10 MHz reference signals. Then data is transmitted through 10-gigabit Ethernet for recording. The system interfaces with an embedded Linux platform on ROACH2, controlled using Katcp protocol over LAN, ensuring precise timing and efficient data handling.}
   \label{Fig3}
   \end{figure}

The architecture of the VLBI data acquisition system, as illustrated in Figure \ref{Fig3}, 
features the VLBI digital backend contained within a black dashed block. 
The ROACH2 platform serves as the backbone of the VLBI digital backend, 
leveraging the ADC module for high-precision analog-to-digital conversions. 
The ADC device operates in a dual-channel input mode, with each channel sampling at 1024MHz, and quantizing the analog input to eight bits per sample.

Underpinned by a high-speed sampling clock, 
the ADC delivers a wideband digital IF signal to the FPGA,
acting as a critical interface between the digital and analog realms. 
In real time, the FPGA processed this wideband digital IF signal, 
converting it into multiple digital baseband signals.
This design employs the PFB digital baseband conversion algorithm contained in the official CASPER library, 
resulting in the output of digital complex baseband data.

%      A figure as large as the width of the column
%-------------------------------------------------------------
%
%      One column rotated figure
%-------------------------------------------------------------

This design utilizes a dual-IF input configuration,
independently processing each digital IF signal  within the FPGA 
to convert them into digital baseband signals. 
The baseband conversion mode is configured as 16x32MHz, 
meaning each IF signal is divided into 16 real-valued digital baseband channels.
With two IF inputs, a total of 32 baseband channels are generated,
each with a bandwidth of 32MHz. 
To refine the output, a channel selection module performs a 32-to-16 selection process to narrow down the signals to sixteen baseband channels, as depicted in Figure \ref{Fig3}. 
Afterward, the selected signals undergo two-bit quantization and
are packaged using the Mark5B data formatter for further processing \citep{Zhang+etal+2013,Guo+etal+2020,Gan+etal+2022}.

Subsequently, the data is transmitted via 10GbE at a rate of 2Gbps and 
is captured and preserved by the Next Generation High-speed Digital Data Record and Playback System (NG-DRAPS). % \citep{Guo+2023}. 
NG-DRAPS not only replicates the core functionalities of Mark6 \citep{Guo+etal+2017}, 
but also introduces advanced features such as playback capabilities and data decoding functions.

\subsection{Digital Baseband Converter Design}

\begin{figure}
   \centering
   \includegraphics[width=\textwidth, angle=0]{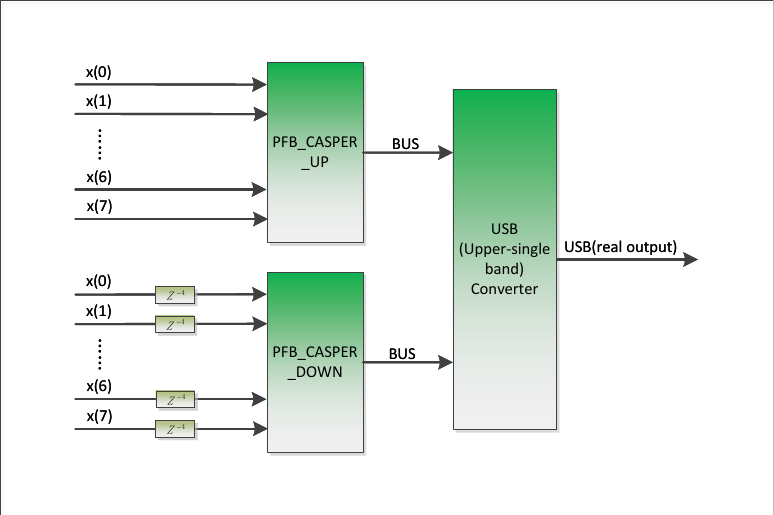}
   \caption{Diagram illustrating the CASPER PFB architecture for both upper and lower band. The $PFB\_CASPER\_UP$ and $PFB\_CASPER\_DOWN$ modules process input signals (x(0)-x(7)) using polyphase decomposition and delay elements $(Z^{-4})$. Outputs are routed through a bus system into the USB (Upper-single band) converter for real output generation.}
   \label{Fig4}
   \end{figure}

The digital baseband conversion algorithm consists of the PFB\_CASPER module \footnote{\url{https://casper.berkeley.edu/wiki/Pfbfirreal}}
\footnote{\url{https://casper.berkeley.edu/wiki/Fftwidebandreal}}
% \citep{pfb1+2019, pfb2+2019}
and SSB module \citep{Lan+etal+2014} module, 
the structure can be seen in Figure \ref{Fig4}. x(0) to x(7) represent the 8 parallel digital signals emerging from the high-speed ADC chip. PFB\_CASPER\_UP and PFB\_CASPER\_DOWN implement the channelization of the wideband IF signal, 
eventually converting the wideband IF signal into multiple digital complex baseband signals with equal bandwidth. 
Both PFB\_CASPER\_UP and PFB\_CASPER\_DOWN can be called from the CASPER library, 
reaffirming its foundation in the PFB algorithm.
%      A figure as large as the width of the column
%-------------------------------------------------------------
%
%      One column rotated figure
%-------------------------------------------------------------

Digital single sideband (SSB) conversion, typically denoted as the upper sideband (USB) and lower sideband (LSB).
The SSB module combines the fundamental modules offered by Xilinx \footnote{\url{https://www.origin.xilinx.com/support/documents/sw_manuals/xilinx14_7/sysgen_gs.pdf}} to perform the upper sideband conversion function on the digital baseband signal. 

%The CASPER library's digital baseband conversion model generates digital complex baseband signals ,
%characterized by an asymmetric spectrum that include both positive and negative frequencies. 
%While real signals are represented by real number,  
%complex signals employ complex numbers to represent both real and imaginary components. 
%To convey equivalent information, a real signal requires a sampling rate twice that of a complex signal, 
%ensuring no data loss. 
%This design combines two digital complex baseband signal structures to increase the sampling rate of the digital baseband signals,
%aligning them with the Nyquist sampling theorem for real signals.
%Consequently, this configuration provides the conditions necessary to convert 
%digital complex baseband signals into digital real baseband signals.

The CASPER library's digital baseband conversion model generates digital complex baseband signals. To obtain real-valued outputs, the complex baseband signal is frequency-shifted by half its bandwidth in both the lower and upper directions, followed by summation of the two shifted components. This process produces a real-valued baseband signal with twice the bandwidth of the original complex baseband signal. By integrating two digital complex baseband signal structures, this design effectively enhances the sampling rate of the digital baseband signals.

For real signal, the original baseband signal can retrieve using either the positive or negative frequency band. 
Despite their conceptual differences,
real and complex signals convey equivalent information. 
To ensure compatibility with CDAS system, the output must consist of digital real baseband signals, 
which calls for the addition of SSB conversion functions based on the digital baseband conversion model in the CASPER library.

As illustrated in Figure \ref{Fig4}, the digital complex baseband signals output by PFB\_CASPER\_UP and PFB\_CASPER\_DOWN exhibit a 180$^\circ$ phase difference.
These two digital complex baseband signals are subsequently converted into digital real baseband signals via the SSB module. 
Importantly, the sampling frequency of this digital real baseband signal is twice its bandwidth.

\subsection{Channel selection and 2-bit quantization design}

In this design, two IF inputs are necessitated with the ADC chip operating in dual-channel mode,
each channel sampling at 1024MHz.
Each IF signal is converted into 16 digital baseband signals, 
resulting to a total of 32 digital baseband signals.

The Mark5B data formatter can encapsulate  digital baseband data of up to 16 channels.
Subsequently, a 32-to-16 selection scheme is employed by the channel selection module to select 16 digital baseband channels from the existing 32 digital baseband signals.
After the channel selection process isfinalized,
the 16 channels of digital baseband data undergo a 2 bits quantization.

VLBI observations typically use 1-bit or 2-bit quantization techniques. 
The 1-bit quantization is often accomplished by direct truncation of the sign bit of the digital baseband signal.
On the other hand, to achieve 2-bit quantization as explained \citep{Zhang+etal+2013} ,
one must apply a Gaussian distribution threshold calculation method to determine the optimal quantization threshold. 
The ideal threshold $H$ needs to satisfy Equation \ref{equ-threshold}, 
where x symbolizes the digital baseband signal and $V_{ref}$ represents the standard deviation of the Gaussian distribution. 
The calculation for deriving the optimal threshold  \citep{Zhang+etal+2013} is specified in Equation \ref{equ-optimal-threshold}.

\begin{equation}
\label{equ-threshold}
\int_{0}^{H} exp(-\frac{x^2}{2V_{ref}^2})\ dx = \int_{-H}^{0} exp(-\frac{x^2}{2V_{ref}^2})\ dx = 0.32
\end{equation}

\begin{equation}
\label{equ-optimal-threshold}
H=0.92V_{ref}
\end{equation}

The variance calculation for a digital baseband signal equates to compute the average power calculation of that digital baseband signal, 
giving that the mean value of the IF signal is 0, 
which means that the value of the standard deviation $V_{ref}$ can be obtained by computing the square root of the average power. 
In an FPGA design, 
assessing the average power of each channel's digital baseband signal yields the variance for each channel, 
which then serves as a basis to calculate the most suitable quantization threshold. 
Real-time update of the estimated optimum quantization threshold is triggered by changes in the input IF signal's power.

\subsection{Mark5B Data Formatting Design}

\begin{figure}[h]
   \centering
   \includegraphics[width=12cm, angle=0]{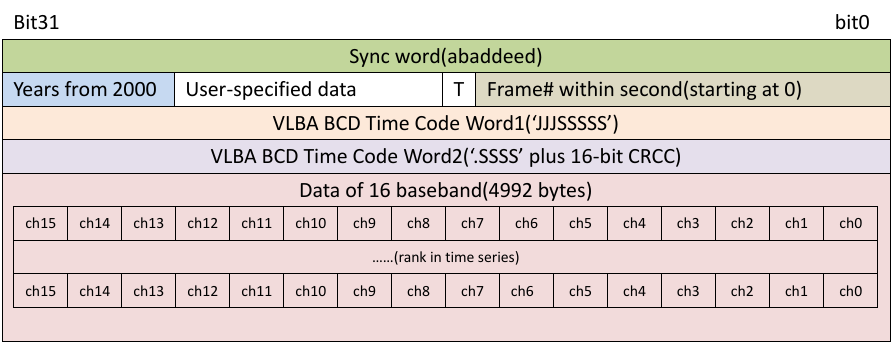}
   \caption{Structure of Mark5B data frame format. The frame begins with a synchronization word (0XABADDEED) followed by metadata fields: years from 2000, user specified data, flag data and the frame number with the second. The data section consists of 16 baseband channels, with channels ordered by time series for processing.}
   \label{Fig-mark5b}
   \end{figure}

The Mark5B data acquisition system utilized in VLBI operations uses a specific data encapsulation process for its 16 channels.
Data from each  channel  is encapsulated in a descending order, 
starting with channel 15 and ending with  channel 0.
The data format of the Mark5B include two primary parts, 
a frame header and a data section \citep{Guo+etal+2020}. 
The frame header occupies 16 bytes, 
contains key structural information necessary for the data part to work properly, and the data section, which is  10000 bytes long, serves as the main repository for the collected data.
The data section is arranged in sequential channel order layout,
following the descending order of channel numbers from ch15, ch14, to ch0.
Each channel, designated as chx, contains 2-bit quantized data.
In this data arrangement,
the amplitude bit takes up the front part while the sign bit occupies the part behind.
%and the Mark5B data formatter frame structure is depicted in Figure 
Figure \ref{Fig-mark5b} provides a visual representation of the structuring of the Mark5B data formatter frame.
This consistent data encapsulation ensures that data from all 16 channels are accurately captured and stored for future analysis in the VLBI system.

   \begin{figure}[h]
   \centering
   \includegraphics[width=12cm, angle=0]{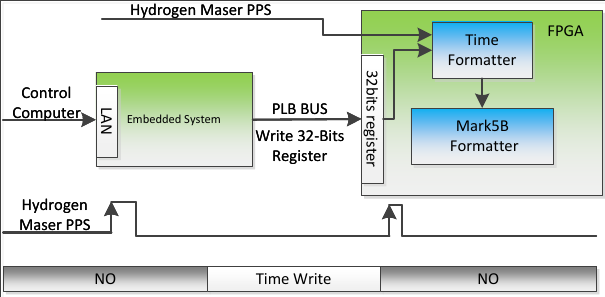}
   \caption{Synchronization framework for the Mark5B time-formatter using a hydrogen maser PPS signal.}
   \label{Fig-sync}
   \end{figure}

The Mark5B data formatter frame header incorporates a time stamp,  facilitated by a Binary-Coded Decimal (BCD) encoding representation of a  truncated Modify Julian Day (MJD).
In this coding system, the final three digits of the day number are symbolized as JJJ, 
while SSSSS represents the current UT second of the data frame. 
The control computer provides the initial time for the time formatter, which is designed in the FPGA. 
For successful operations, synchronization must occur between the control computer and the Network Time Protocol (NTP) server.

The control computer writes the initial time into the software register of the FPGA via the KATCP protocol and waits for the hydrogen clock's pulse per second (PPS)  signal to trigger the synchronization. 
The time synchronization technique uses the rising edge of the hydrogen clock PPS to trigger time synchronization and formatter initialization. 
The time formatter synchronization mechanism is shown in Figure \ref{Fig-sync}.
Overall, the Mark5B data formatter ensures the high precision and accuracy of data recording and synchronization necessary for effective VLBI operations.

\subsection{Brief summary}

This design, built on the ROACH2 astronomical digital platform, 
implements a VLBI digital backend.
The backbone of this design is dependent on the polyphase filter bank,
a key algorithm obtained from the official CASPER library.
Primarily intended for use in geodetic VLBI observation, the VLBI digital backend is configured to receive two intermediate frequency signals,
each with a bandwidth of 512MHz.
Following the necessary processing, 
the backend then outputs 16 digital baseband signals,
each with a bandwidth of 32MHz.

%
%               one-column-spanning table
%________________________________________ Table 1: Use_of_the routines
\begin{table}
\begin{center}
\caption{ The Specification of VLBI Digital Backend.}\label{tab-vlbi}

%%Please Capitalize the First Letter of Each Notional Word in table's caption

 \begin{tabular}{clcl}
  \hline\noalign{\smallskip}
No &  Technology      & Parameters                     \\
  \hline\noalign{\smallskip}
1  & IF Bandwidth & 512MHz \\
2  & Number of Channels & 16 \\
3  & Baseband Bandwidth & 32MHz \\
4  & Number of IF Inputs & 2 \\
5  & Signal Output Form & Real Signal \\
6  & Passband/Stopband Factor & 0.875 \\
7  & Data Formatter & Mark5B \\
8  & Quantization Mode & 2-bit \\
9  & Data Rate & 2Gbps \\
10  & Number of Baseband Outputs & 32-to-16 \\
  \noalign{\smallskip}\hline
\end{tabular}
\end{center}
\end{table}

This system, like the traditional VLBI system, 
also generates digital real baseband signals,
with the data formatter adhering to the Mark5B standard.
In this configuration, each IF operates with a sampling frequency of 1024MHz and an observation bandwidth of 512MHz. The data output rates come up to 2Gbps, underpinned by a 10 Gigabit Ethernet (10GbE) connection.
For a more comprehensive picture,
The backend specifications are summarized in Table \ref{tab-vlbi},
highlighting the technical parameters of the VLBI digital backend.

\section{Multi-Station VLBI Observation Experiments}

Observation experiments of long-baseline interferometry were carried out on the VLBI digital backend based on the ROACH2 platform. 
The Tianma 65m radio telescope, the Urumqi 25m radio telescope, and the Kunming 40m radio telescope all took part in the experiment.

At the Tianma 65m radio telescope station, 
both the CDAS2-D terminal used in the Chang'e Lunar Exploration Program 
and the ROACH2-based VLBI digital backend were installed. 
The Urumqi 25m station and the Kunming 40m station both made use of the CDAS2-D terminal.
The observed radio source was NRAO530, with observation periods spanning 30 minutes. 
Dual-IF observation was carried out simultaneously using X-band and S-band. 
The first thirteen channels were X-band and the last three channels were S-band. 
The observation frequency settings are shown in Table \ref{tab-setting}.

\begin{table}[h]
\center
\caption{The Setting of Observing Frequencies and Channels}
\label{tab-setting}
\begin{tabular}{|c|cllllcll|}
\hline
\multirow{2}{*}{Channel}                                         & \multicolumn{8}{c|}{X-band}                                                                                                                                                                                                 \\ \cline{2-9} 
                                                                 & \multicolumn{1}{c|}{1}    & \multicolumn{1}{c|}{2}    & \multicolumn{1}{c|}{3}    & \multicolumn{1}{c|}{4}    & \multicolumn{1}{c|}{5}    & \multicolumn{1}{c|}{6}    & \multicolumn{1}{c|}{7}    & \multicolumn{1}{c|}{8}  \\ \hline
\begin{tabular}[c]{@{}c@{}}Center Frequency\\ (MHz)\end{tabular} & \multicolumn{1}{l|}{8180} & \multicolumn{1}{l|}{8212} & \multicolumn{1}{l|}{8244} & \multicolumn{1}{l|}{8276} & \multicolumn{1}{l|}{8308} & \multicolumn{1}{l|}{8340} & \multicolumn{1}{l|}{8372} & 8404                    \\ \hline
\multirow{2}{*}{Channel}                                         & \multicolumn{5}{c|}{X-band}                                                                                                               & \multicolumn{3}{c|}{S-band}                                                     \\ \cline{2-9} 
                                                                 & \multicolumn{1}{c|}{9}    & \multicolumn{1}{c|}{10}   & \multicolumn{1}{c|}{11}   & \multicolumn{1}{c|}{12}   & \multicolumn{1}{c|}{13}   & \multicolumn{1}{c|}{14}   & \multicolumn{1}{c|}{15}   & \multicolumn{1}{c|}{16} \\ \hline
\begin{tabular}[c]{@{}c@{}}Center Frequency\\ (MHz)\end{tabular} & \multicolumn{1}{l|}{8436} & \multicolumn{1}{l|}{8468} & \multicolumn{1}{l|}{8500} & \multicolumn{1}{l|}{8532} & \multicolumn{1}{l|}{8564} & \multicolumn{1}{l|}{2208} & \multicolumn{1}{l|}{2240} & 2272                    \\ \hline
\end{tabular}
\end{table}

\begin{figure}[h]
   \centering
   \includegraphics[width=12cm, angle=0]{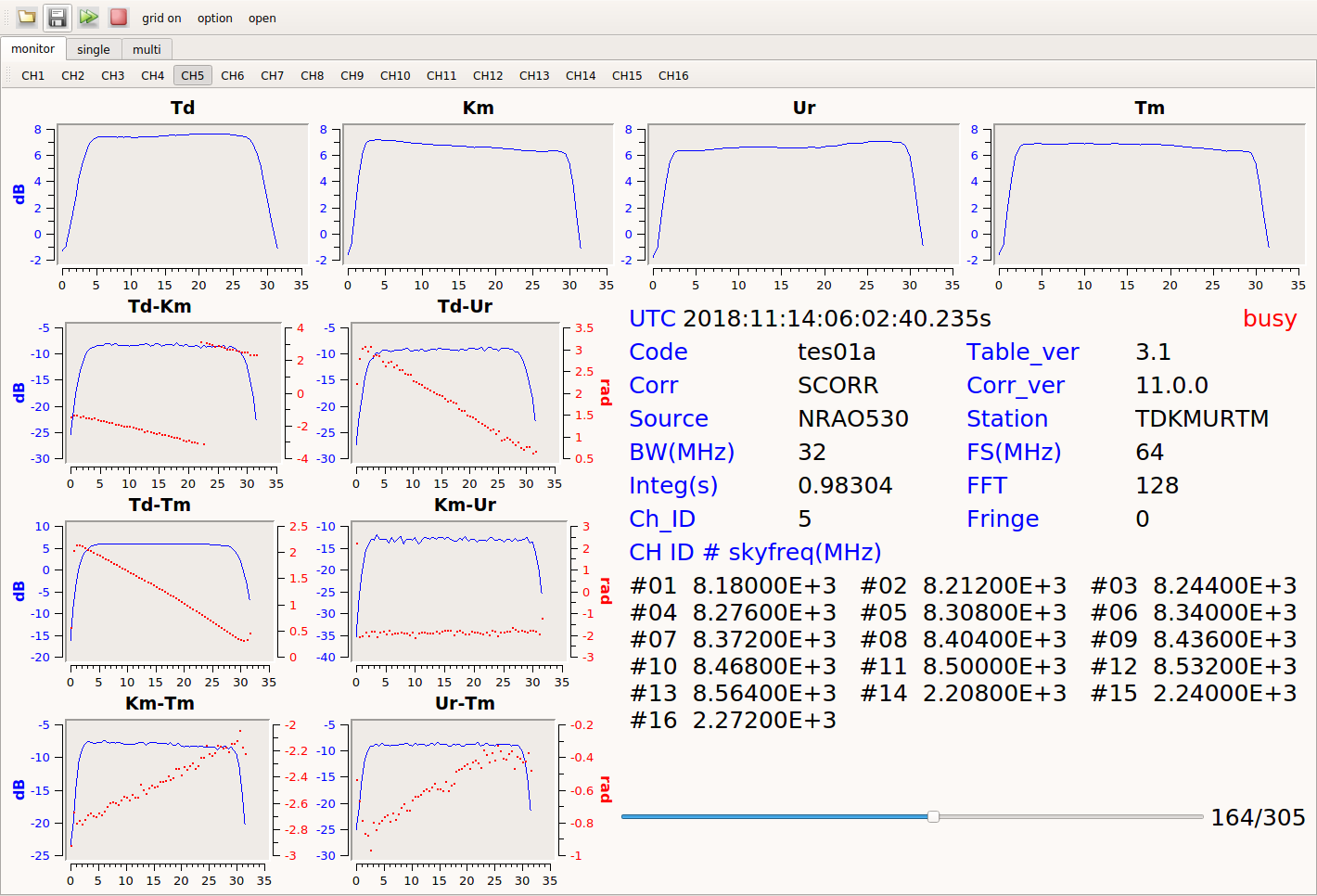}
   \caption{Correlation analysis interface for the observation of the source NRAO530 in 5th Channel (X Band). The top panel shows the spectral amplitude for individual stations over frequency channels, while the lower panels display cross-correlation results between station pairs (e.g., Td-Km, Td-Ur, etc.), with amplitude (blue) and phase (red).}
   \label{Fig-scorr-5th}
   \end{figure}

\begin{figure}[h]
   \centering
   \includegraphics[width=12cm, angle=0]{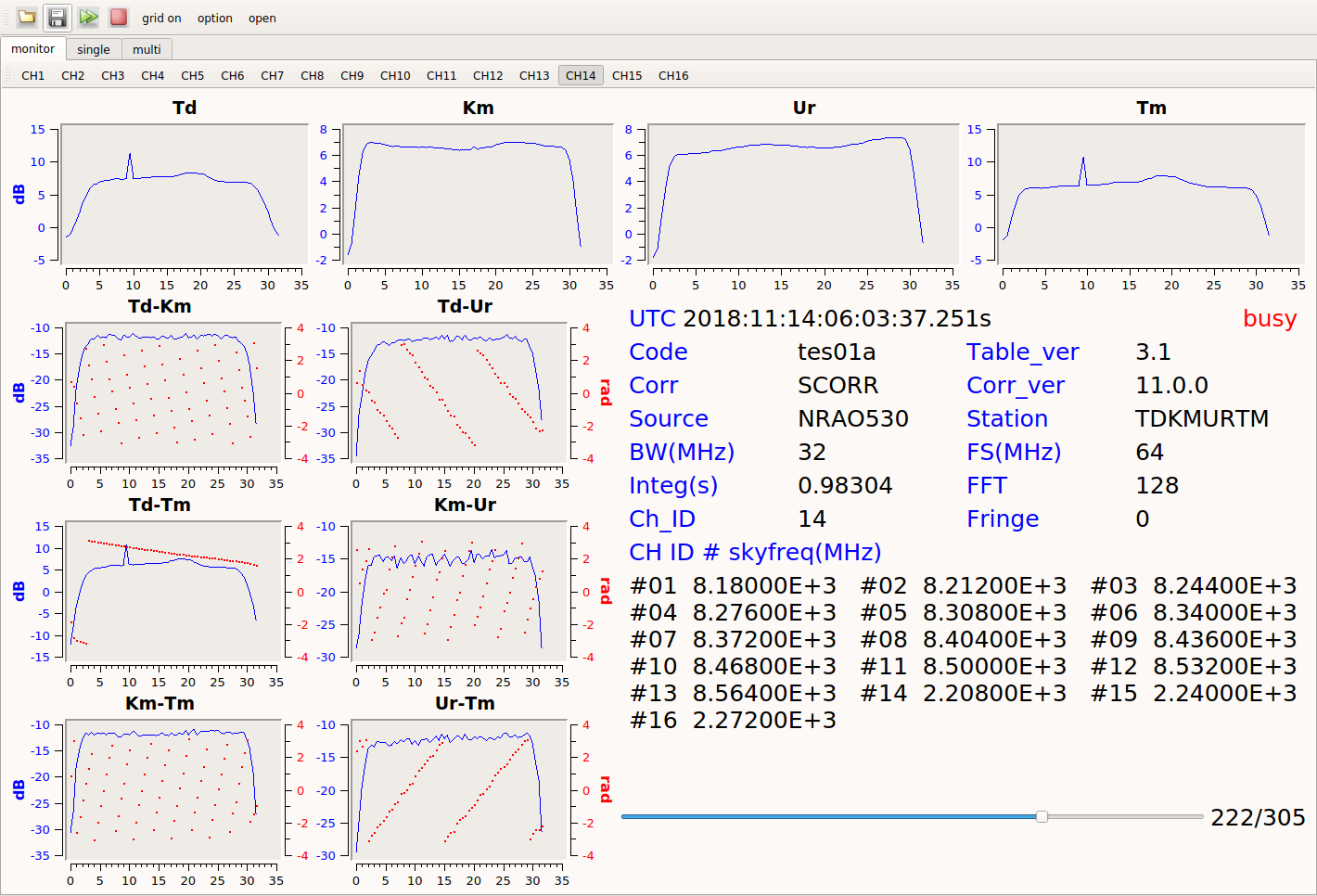}
   \caption{Correlation analysis interface for the observation of the source NRAO530 in  14th Channel (S Band).}
   \label{Fig-scorr-14th}
   \end{figure}

The cross-correlation data between multiple stations was processed using the VLBI correlator located at the Shanghai Astronomical Observatory.
The processed results of channel five and fourteen can be viewed in Figures \ref{Fig-scorr-5th} and Figure \ref{Fig-scorr-14th}. 
In these figures, the CDAS2-D terminals at Tianma, Kunming, and Nanshan stations are denoted by the abbreviations Tm, Km, and Ur respectively.
Additionally, the term Td stands for the VLBI digital backend based on the ROACH2 hardware platform.

The auto-correlation amplitude spectra of Td, Tm, Km, and Ur are shown in the first row of amplitude spectra in Figures \ref{Fig-scorr-5th} and Figures \ref{Fig-scorr-14th}, respectively. 
These visuals provide insight into the performance and correlation between their units.
The remaining six amplitude spectra, which illustrate cross-correlation amplitude and phase spectra, 
provide crucial data on the cross-correlation findings across the six baseline connections between Td, Tm, Km, and Ur. 
The color coding used red to denote the phase spectrum, 
and blue for the amplitude spectrum. 
Based on the processing results obtained from the correlator at Shanghai Observatory correlator center,
fringes were successfully captured across all 16 channels.
This confirms the successful operation of the VLBI digital backend based on the ROACH2 platform,
indicating that it was capable of reliable data capture and processing during these long baseline interferometry experiments.

\section{Conclusion and Future Prospects}

In conclusion, the ROACH2 platform-based VLBI digital backend has successfully demonstrated its capability to support dual-IF analog signal input with a bandwidth of 512MHz per input signal. 
After digitization, each IF signal generates sixteen digital baseband signals,
each with a bandwidth of 32MHz.
Depending on the requirements of the observation,
any sixteen digital baseband signals can be selected for output.
Furthermore, the adoption of the Mark5B standard data formatter ensures
a high level of compatibility and seamless integration with existing VLBI infrastructures.

The experimental results validate the system's conventional VLBI observation capabilities.
The successful detection of correlation fringes across all channels during multi-station VLBI experiments demonstrates the backend's robust performance in processing wideband signals and its compatibility with existing VLBI infrastructure.
Additionally, Xinjiang Astronomical Observatory has extended the platform's versatility by developing a pulsar digital backend in incoherent mode. 
This pulsar backend has undergone extensive validation through multiple observation campaigns, successfully producing high-quality pulse profiles through long-term integration processing, further validating the platform's signal processing capabilities.
These successful implementations demonstrate the ROACH2 platform's versatility in supporting diverse astronomical applications, from high-precision VLBI observations to specialized pulsar studies, including both single-pulse detection and real-time analysis capabilities.

The mature development ecosystem of the ROACH2 platform, featuring comprehensive user interfaces and extensive documentation, significantly streamlines the testing and verification processes while substantially reducing development costs and implementation time.
The platform's integration with the CASPER library's advanced signal processing algorithms dramatically accelerates development cycles by offering well-tested and optimized implementations of complex digital signal processing functions, enabling rapid deployment of sophisticated astronomical observation systems.

Looking forward, the ROACH2 platform continues to offer significant potential for advancing radio astronomy digital backend development, particularly in areas requiring flexible, cost-effective solutions.
A key area of focus is the development of modular designs that enhance flexibility and scalability,
allowing for rapid customization to meet diverse observational need.
While the current implementation uses the Mark5B data format, 
the system can be readily adapted to support the VLBI Data Interchange Format (VDIF) 
with an estimated development time of approximately two months, 
demonstrating the platform's adaptability to evolving VLBI standards.
By leveraging the platform's versatility, researchers can develop multi-functional backends that support various applications, including pulsar, VLBI, and spectral line observations, on a unified platform.
As VLBI technology continues to evolve, emphasis on higher data rates, wider bandwidths, and more efficient data processing pipelines will be critical. The ROACH2 platform provides a solid foundation for addressing these challenges, offering a cost-effective and reliable solution for future advancements in astronomical instrumentation.

In conclusion, our implementation of a VLBI digital backend on the ROACH2 platform demonstrates not only the system's capability to meet current observational requirements but also its potential to facilitate innovative developments in next-generation astronomical instrumentation through its flexible architecture and extensive development resources.

\begin{acknowledgments}
We would like to express our gratitude to the anonymous reviewer for their valuable comments and suggestions. The authors would like to thank Wasim Raja for his assistance in polishing the language of this manuscript.
This work was funded by National SKA Program of China under No.2020SKA0110300, 
and supported by the Astrometric Reference Frame project, Grant No. JZZX-0101.
Jiyun was funded by the National Natural Science Foundation of China (NSFC) under No.12103079.
Shaoguang was supported by Youth Innovation Promotion Association CAS Program under No.2021258 and International Partnership Program of the Chinese Academy of Sciences Grant No. 018GJHZ2024025GC.
\end{acknowledgments}


\vspace{5mm}
\facilities{CDAS, FPGA}


\software{astropy \citep{2013A&A...558A..33A,2018AJ....156..123A},  
         matplotlib \citep{Hunter:2007},
         PyQt
          }



%\appendix


\bibliography{ref}{}

\begin{thebibliography}{}
\expandafter\ifx\csname natexlab\endcsname\relax\def\natexlab#1{#1}\fi
\providecommand{\url}[1]{\href{#1}{#1}}
\providecommand{\dodoi}[1]{doi:~\href{http://doi.org/#1}{\nolinkurl{#1}}}
\providecommand{\doeprint}[1]{\href{http://ascl.net/#1}{\nolinkurl{http://ascl.net/#1}}}
\providecommand{\doarXiv}[1]{\href{https://arxiv.org/abs/#1}{\nolinkurl{https://arxiv.org/abs/#1}}}

\bibitem[{{Astropy Collaboration} {et~al.}(2013){Astropy Collaboration},
  {Robitaille}, {Tollerud}, {Greenfield}, {Droettboom}, {Bray}, {Aldcroft},
  {Davis}, {Ginsburg}, {Price-Whelan}, {Kerzendorf}, {Conley}, {Crighton},
  {Barbary}, {Muna}, {Ferguson}, {Grollier}, {Parikh}, {Nair}, {Unther},
  {Deil}, {Woillez}, {Conseil}, {Kramer}, {Turner}, {Singer}, {Fox}, {Weaver},
  {Zabalza}, {Edwards}, {Azalee Bostroem}, {Burke}, {Casey}, {Crawford},
  {Dencheva}, {Ely}, {Jenness}, {Labrie}, {Lim}, {Pierfederici}, {Pontzen},
  {Ptak}, {Refsdal}, {Servillat}, \& {Streicher}}]{2013A&A...558A..33A}
{Astropy Collaboration}, {Robitaille}, T.~P., {Tollerud}, E.~J., {et~al.} 2013,
  \aap, 558, A33, \dodoi{10.1051/0004-6361/201322068}

\bibitem[{{Astropy Collaboration} {et~al.}(2018){Astropy Collaboration},
  {Price-Whelan}, {Sip{\H{o}}cz}, {G{\"u}nther}, {Lim}, {Crawford}, {Conseil},
  {Shupe}, {Craig}, {Dencheva}, {Ginsburg}, {VanderPlas}, {Bradley},
  {P{\'e}rez-Su{\'a}rez}, {de Val-Borro}, {Aldcroft}, {Cruz}, {Robitaille},
  {Tollerud}, {Ardelean}, {Babej}, {Bach}, {Bachetti}, {Bakanov}, {Bamford},
  {Barentsen}, {Barmby}, {Baumbach}, {Berry}, {Biscani}, {Boquien}, {Bostroem},
  {Bouma}, {Brammer}, {Bray}, {Breytenbach}, {Buddelmeijer}, {Burke},
  {Calderone}, {Cano Rodr{\'\i}guez}, {Cara}, {Cardoso}, {Cheedella}, {Copin},
  {Corrales}, {Crichton}, {D'Avella}, {Deil}, {Depagne}, {Dietrich}, {Donath},
  {Droettboom}, {Earl}, {Erben}, {Fabbro}, {Ferreira}, {Finethy}, {Fox},
  {Garrison}, {Gibbons}, {Goldstein}, {Gommers}, {Greco}, {Greenfield},
  {Groener}, {Grollier}, {Hagen}, {Hirst}, {Homeier}, {Horton}, {Hosseinzadeh},
  {Hu}, {Hunkeler}, {Ivezi{\'c}}, {Jain}, {Jenness}, {Kanarek}, {Kendrew},
  {Kern}, {Kerzendorf}, {Khvalko}, {King}, {Kirkby}, {Kulkarni}, {Kumar},
  {Lee}, {Lenz}, {Littlefair}, {Ma}, {Macleod}, {Mastropietro}, {McCully},
  {Montagnac}, {Morris}, {Mueller}, {Mumford}, {Muna}, {Murphy}, {Nelson},
  {Nguyen}, {Ninan}, {N{\"o}the}, {Ogaz}, {Oh}, {Parejko}, {Parley}, {Pascual},
  {Patil}, {Patil}, {Plunkett}, {Prochaska}, {Rastogi}, {Reddy Janga},
  {Sabater}, {Sakurikar}, {Seifert}, {Sherbert}, {Sherwood-Taylor}, {Shih},
  {Sick}, {Silbiger}, {Singanamalla}, {Singer}, {Sladen}, {Sooley},
  {Sornarajah}, {Streicher}, {Teuben}, {Thomas}, {Tremblay}, {Turner},
  {Terr{\'o}n}, {van Kerkwijk}, {de la Vega}, {Watkins}, {Weaver}, {Whitmore},
  {Woillez}, {Zabalza}, \& {Astropy Contributors}}]{2018AJ....156..123A}
{Astropy Collaboration}, {Price-Whelan}, A.~M., {Sip{\H{o}}cz}, B.~M., {et~al.}
  2018, \aj, 156, 123, \dodoi{10.3847/1538-3881/aabc4f}

\bibitem[{Baudry(2008)}]{Baudry+2008}
Baudry, A. 2008, in 2nd MCCT-SKADS, 002

\bibitem[{Chen {et~al.}(2014)Chen, Li, \& Zhang}]{Lan+etal+2014}
Chen, L., Li, J.~Y., \& Zhang, Q. e.~a. 2014, 1098

\bibitem[{Dewdney {et~al.}(2009)Dewdney, Hall, \&
  Schilizzi}]{Dewdney+etal+2009}
Dewdney, P.~E., Hall, P.~J., \& Schilizzi, R. T. e.~a. 2009, Proceeding of the
  IEEE, 97, 1482

\bibitem[{Gan {et~al.}(2022)Gan, Guo, \& He}]{Gan+etal+2022}
Gan, J.~Y., Guo, S.~G., \& He, X. e.~a. 2022, Chinese Space Science and
  Technology, 42, 46

\bibitem[{Guo {et~al.}(2020)Guo, Li, \& Zhu}]{Guo+etal+2020}
Guo, S.~G., Li, J.~Y., \& Zhu, R. J. e.~a. 2020, ACTA ASTRONOMICA SINICA, 3, 66

\bibitem[{Guo {et~al.}(2017)Guo, Zhu, \& Zheng}]{Guo+etal+2017}
Guo, S.~G., Zhu, R.~J., \& Zheng, W. M. e.~a. 2017, Astronomical Research and
  Technology, 14, 281

\bibitem[{Hunter(2007)}]{Hunter:2007}
Hunter, J.~D. 2007, Computing in Science \& Engineering, 9, 90,
  \dodoi{10.1109/MCSE.2007.55}

\bibitem[{Jiang {et~al.}(2024)Jiang, Chen, Gan, Sun, Zhu, Li, Zhu, Wu, Chen,
  Zhang, \& An}]{jiangpeng-fast-core-array}
Jiang, P., Chen, R., Gan, H., {et~al.} 2024, Astronomical Techniques and
  Instruments, 1, 84, \dodoi{10.61977/ati2024012}

\bibitem[{Li {et~al.}(2019)Li, Wang, Wei, \& Lin}]{Li+etal+2019}
Li, C., Wang, C., Wei, Y., \& Lin, Y. 2019, Science, 365, 238

\bibitem[{Li {et~al.}(2018)Li, Wang, \& Qian}]{Li+etal+2018}
Li, D., Wang, P., \& Qian, L. e.~a. 2018, IEEE Microwave Magazine, 19, 112

\bibitem[{Liu {et~al.}(2024)Liu, Shen, Hong, Ye, Li, Wang, Zhao, Fu, Zhong,
  Wang, {et~al.}}]{liu2024tianma}
Liu, Q., Shen, Z., Hong, X., {et~al.} 2024, Astronomical Techniques and
  Instruments, 1, 239

\bibitem[{Nan {et~al.}(2011)Nan, Li, \& Jin}]{Nan+etal+2011}
Nan, R.~D., Li, D., \& Jin, C. J. e.~a. 2011, International Journal of Modern
  Physics D, 20, 989

\bibitem[{Niell {et~al.}(2010)Niell, Bark, \& Beaudoin}]{Niell+etal+2010}
Niell, A., Bark, M., \& Beaudoin, C. e.~a. 2010, L396

\bibitem[{Pei {et~al.}(2017)Pei, Li, \& Yuan}]{Pei+etal+2017}
Pei, X., Li, J., \& Yuan, J. P. e.~a. 2017, PROGRESS IN ASTRONOMY, 35, 244

\bibitem[{Takefuji {et~al.}(2010)Takefuji, Takeuchi, \&
  Tsutsumi}]{Takefuji+etal+2010}
Takefuji, K., Takeuchi, K., \& Tsutsumi, M. e.~a. 2010, L378

\bibitem[{Thompson {et~al.}(2017)Thompson, James, \&
  Gorge}]{Thompson+etal+2017}
Thompson, A.~R., James, M.~M., \& Gorge, W. e.~a. 2017, Interferometry and
  Synthesis in Radio Astronomy, 3rd edn. (Switzerland: Springer Nature)

\bibitem[{Tuccari {et~al.}(2012)Tuccari, Buttacio, \& Alef}]{Tuccari+etal+2012}
Tuccari, G., Buttacio, S., \& Alef, W. e.~a. 2012

\bibitem[{{Wang} {et~al.}(2023){Wang}, {Xu}, {Ma}, {Liu}, {Liu}, {Zhang},
  {Pei}, {Chen}, {Manchester}, {Lee}, {Zheng}, {K{\"a}rcher}, {Zhao}, {Li},
  {Li}, {S{\"u}ss}, {Reichert}, {Zhu}, {Wang}, {Li}, {Li}, {Li}, {Kazezkhan},
  {Yan}, {Wu}, {Cui}, {Zhang}, \& {Li}}]{wangna-qtt}
{Wang}, N., {Xu}, Q., {Ma}, J., {et~al.} 2023, Science China Physics,
  Mechanics, and Astronomy, 66, 289512, \dodoi{10.1007/s11433-023-2131-1}

\bibitem[{Wu {et~al.}(2017)Wu, Liu, \& Li}]{Wu+etal+2017}
Wu, Y.~J., Liu, Q.~H., \& Li, J. e.~a. 2017, Astronomical Research and
  Technology, 14, 1

\bibitem[{Xiang(2005)}]{Xiang+2005}
Xiang, Y. 2005, PhD thesis, Shanghai Astronomical Observatory, Chinese Academy
  of Sciences

\bibitem[{Xu {et~al.}(2015)Xu, Li, \& Zhang}]{Xu+etal+2015}
Xu, Y.~H., Li, J.~Y., \& Zhang, Y. Q. e.~a. 2015, Astronomical Research and
  Technology, 12, 480

\bibitem[{Zhang {et~al.}(2013)Zhang, Wu, \& Yu}]{Zhang+etal+2013}
Zhang, B.~J., Wu, Y.~J., \& Yu, W. e.~a. 2013, Astronomical Research and
  Technology, 10, 219

\bibitem[{Zhu {et~al.}(2018)Zhu, Wu, \& Li}]{Zhu+etal+2016}
Zhu, R.~J., Wu, Y.~J., \& Li, J. Y. e.~a. 2018, L1130

\end{thebibliography}
\bibliographystyle{aasjournal}



\end{CJK*}
\end{document}